\documentclass[aps,prl,superscriptaddress,notitlepage,nopacs,amsmath,amstex,amssymb,citeautoscript,longbibliography,floatfix,letter,twocolumn]{revtex4-2}
\usepackage{graphicx,amsmath,amssymb,amsfonts,color,physics}
\usepackage{nicefrac}
\usepackage{multirow, makecell, ragged2e}
\usepackage[titletoc,title]{appendix}
\usepackage{bm}
\usepackage[normalem]{ulem}
\usepackage{float,lipsum}
\usepackage{algpseudocode}
\usepackage{algorithm,soul}
\usepackage{mathtools}
\usepackage{graphicx}
\usepackage{subfigure}
\usepackage{adjustbox}
\usepackage{bm}
\usepackage{color}
\usepackage{braket}
\usepackage{standalone}
\usepackage{multirow}
\usepackage{tikz}
\usepackage{mathrsfs}
\usepackage{dsfont}
\usepackage{comment}
\usepackage{amsmath}
\usepackage[colorlinks,bookmarks=true,citecolor=blue,linkcolor=red,urlcolor=blue]{hyperref}
\usepackage{cleveref}
\usepackage{orcidlink}

\usepackage{graphicx,amsmath,amssymb,amsfonts, color}
\usepackage{nicefrac}
\usepackage{multirow, makecell, ragged2e}
\usepackage[titletoc,title]{appendix}
\usepackage{bm}
\usepackage{float,lipsum}
\usepackage{algpseudocode}
\usepackage{algorithm}
\renewcommand{\vec}[1]{\mbox{\boldmath$#1$}}

\newcommand{\be}{\begin{equation}}
\newcommand{\ee}{\end{equation}}

\renewcommand{\vec}[1]{\bm{#1}}

\begin{document}

\title{BCS Stripe Phase in Coupled Bilayer Superconductors}
\author{Uddalok Nag}
\affiliation{Department of Physics, 104 Davey Lab, The Pennsylvania State University, University Park, Pennsylvania 16802, USA}
\author{Jonathan Schirmer}
\affiliation{Department of Physics, 104 Davey Lab, The Pennsylvania State University, University Park, Pennsylvania 16802, USA}
\affiliation{Department of Physics, William \& Mary, Williamsburg, Virginia 23187, USA}
\author{Enrico Rossi}
\affiliation{Department of Physics, William \& Mary, Williamsburg, Virginia 23187, USA}
\author{C.-X. Liu}
\affiliation{Department of Physics, 104 Davey Lab, The Pennsylvania State University, University Park, Pennsylvania 16802, USA}
\author{J. K. Jain\orcidlink{000-0003-0082-5881}}
\affiliation{Department of Physics, 104 Davey Lab, The Pennsylvania State University, University Park, Pennsylvania 16802, USA}

\begin{abstract}
As a signature of competing correlations, stripes occur in a variety of strongly correlated systems, such as high temperature superconductors (SCs) and quantum Hall effect. We study a double layer SC in the presence of a parallel magnetic field $B$ within the Bogoliubov-de Gennes framework. 
We find that for low $B$ the system remains in the ``Bardeen-Cooper-Schrieffer (BCS) phase" with a spatially uniform gap, but with increasing $B$, a transition occurs into a phase which contains stripes of the BCS phase separated by regions where the interlayer phase difference rotates by $2\pi$ due to the presence of inter-layer vortices. This stripe phase is predicted to manifest through oscillations in the amplitude of the SC gap and an alternating pattern of supercurrents. We will comment on the relation to previous works based on the Landau-Ginzburg formalism as well as on the possible experimental realization and signature of this phase.
\end{abstract}
\graphicspath{{./Figures/}}

\maketitle

{\it Introduction -} Stripes occur in many strongly correlated systems and are a signature of competing orders. In the quantum Hall effect they occur at a filling of the form $\nu=n+1/2$ where the $\nu=n$ and $\nu=n+1$ phases compete~\cite{Lilly99,Fogler96}, and in high temperature superconductors (SCs) they arise because the holes in an antiferromagnetic background want to cluster together but a competing phase separation has a prohibitive Coulomb energy cost~\cite{Emery99}. Here we predict that in a bilayer SC in the presence of a  magnetic field parallel to the layers, a state consisting of  stripes of the Bardeen-Cooper-Schrieffer (BCS) phase is stabilized in some parameter range.

Liu considered a bilayer Ising SC model in the presence of a parallel magnetic field using the Ginzburg-Landau (GL) approach~\cite{Liu17}, which is valid near the critical temperature. With increasing magnetic field, a phase transition was predicted from the BCS phase into a phase with the gap function $\Delta_{\pm}(x)=|\Delta|e^{\pm iqx}$, where $\pm$ labels two SC layers. As $|\Delta|$ is spatially uniform and each SC layer has a single $q$ value, we refer to this phase as the ``layer helical phase" below. 
We note that the layer helical phase is also an example of the Fulde-Ferrell-Larkin-Ovchinnikov (FFLO) phase~\cite{Fulde64,larkin1965nonuniform,agterberg2012magnetoelectric,kinnunen2018fulde}. In a system with time reversal and spatial inversion symmetry, which ensures that for every electron with the momentum $\vec{k}$ on the Fermi surface there exists another with the momentum $\vec{-k}$ and opposite spin, Cooper pairs with zero momentum are stabilized, as assumed in the BCS theory. But if any of these symmetries are broken, for example the time reversal symmetry (due to 
a magnetic field) or spatial inversion symmetry (due to antisymmetric spin-orbit coupling), the resulting Cooper pairs can have non-zero center of mass momentum $\vec{q}$, producing the FFLO superconducting phase described by a gap parameter of the form $\Delta(\vec{r})=|\Delta(\vec{r})|e^{i\vec{q}\cdot\vec{r}}$. 
Even though the FFLO phase was predicted six decades ago, its realization has been challenging and it has been reported in very few condensed matter~\cite{Bianchi2003,Matsuda2007} and atomic systems~\cite{Partridge2006,Parish2007,Kinnunen2018}.

\begin{figure*}[t]
        \includegraphics[width = \textwidth]{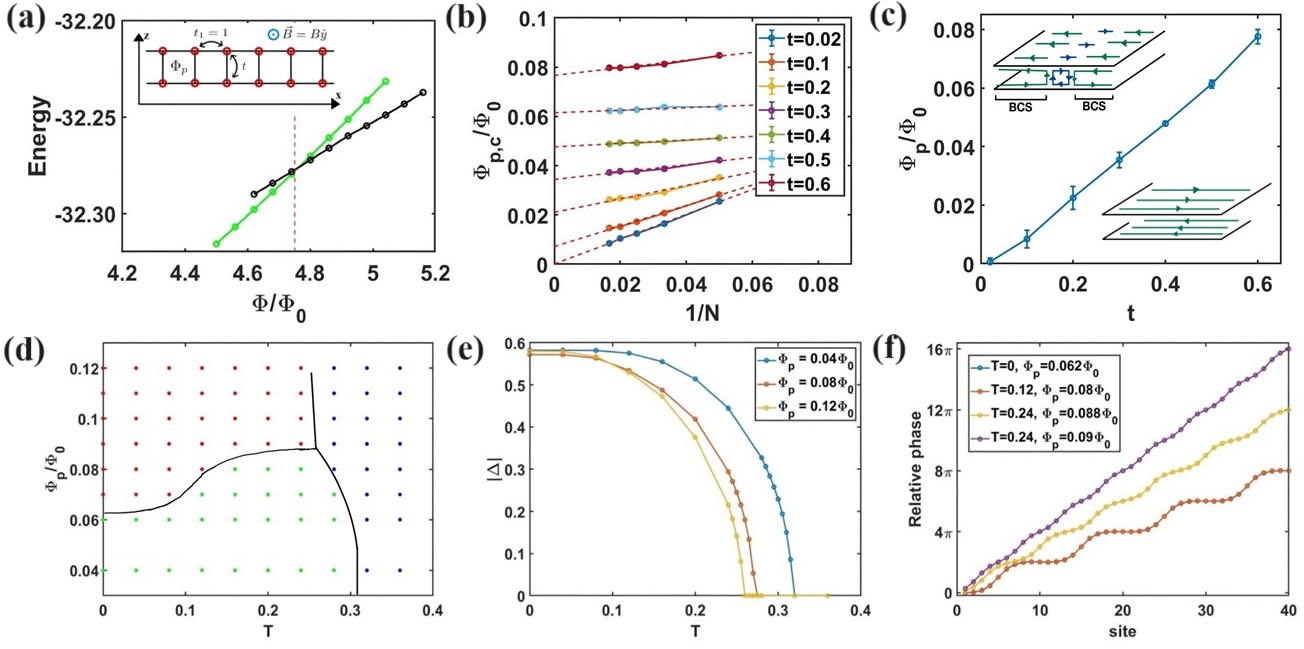}
%\vspace{-2cm}
\caption{(a) Transition from BCS phase (green) to stripe phase (black) as a function of $\Phi/\Phi_0$ for $U=4$, $t=0.6$, $\mu=-3$, $N=60$, and $T=0$. In the vicinity of the transition we find two self-consistent solutions. The inset shows the side view of the lattice model. (b) Determination of the thermodynamic limit of for the critical flux per plaquette, $\Phi_{p,c}$, for several values of $t$. 
(c) $B-t$ phase diagram at $T=0$, $U=4$, and $\mu = -3$. The error in the thermodynamic value of the critical flux is determined by performing linear fits, as in (b), with different numbers of points. The insets show schematics of the current flow patterns for the two phases. (d) The $B-T$ phase diagram for $U=4$, $\mu=-3$, $N=40$, and $t=0.5$. The green, red and blue dots show BCS, stripe and normal phases. (e) The average gap amplitude as a function of temperature for three different values of $\Phi_p/\Phi_0$. (f) The relative phase at 4 different points in the phase diagram \ref{fig:phasediagram}(d) as we first move along the transition line from BCS to stripe phase and then along the BCS to normal phase. The plots for $T=0$, $\Phi_p = 0.062\Phi_0$ and $T=0.12$, $\Phi_p=0.08\Phi_0$ coincide.}
    \label{fig:phasediagram}
\end{figure*}

Indirect evidence for the phase transition predicted in Ref.~\cite{Liu17} has been observed in bilayer or multi-layer Ising SCs~\cite{Zhao2023centrosym,Wan23}, characterized by an upturn in the in-plane upper critical magnetic field when lowering temperature, as well as by broken translational and rotational symmetries~\cite{Croitoru17}. Given the experimental feasibility of this remarkable physics, we theoretically evaluate the phase diagram of a bilayer SC in the presence of an in-plane $B$ by obtaining self-consistent solutions of the Bogoliubov-de Gennes (BdG) equations for electrons on a lattice subject to an attractive Hubbard $U$ interaction. The BdG formalism, which is valid for a wider range of parameters than the GL approach and also allows for spatial variation of the complex gap function, reveals striking physics beyond that in Ref.~\cite{Liu17}. Most remarkably, as the magnetic field is increased, a transition takes place into a phase that consists of stripes of the BCS phase separated by inter-layer vortices (namely $2\pi$ phase winding of the order parameter between two layers), as shown schematically in the inset of Fig.~\ref{fig:phasediagram}(c). This  ``BCS stripe phase" is an example of a highly nonuniform FFLO state. An immediately testable prediction of our work is that the phase occurring close to the transition line in the experiment of Ref.~\cite{Wan23} is a BCS stripe phase, the striped character of which can be revealed by a measurement of the local order parameter, whose amplitude should show periodic oscillations in the direction perpendicular to the magnetic field, as well as by the distribution of local supercurrent. This phase continuously evolves into the layer helical phase of Ref.~\cite{Liu17} in the limit of large $B$ where the layers become effectively decoupled.
An observation of the stripe phase through substantial variation in the gap amplitude will, we believe, provide a more direct confirmation of the underlying FFLO physics.

{\it Model Hamiltonian and Method -}  We model each layer of the double-layer SC as a square lattice, with intra-layer hopping $t_1=1$  and an inter-layer hopping $t$ (inset of Fig.~\ref{fig:phasediagram}(a)).  Superconductivity is incorporated through an on-site attractive interaction term, i.e. a negative-$U$ Hubbard model. The Hamiltonian is given by $\mathcal{H}= \mathcal{H}_0 +\mathcal{H}_I $ with
\begin{eqnarray} 
%\begin{equation}
\label{hamiltonian}
%    \begin{split}
       && \mathcal{H}_0 =-\sum_{\hat{\delta}, l, {\vec{j}}, \sigma} \Big(e^{iA_{\hat{\delta}, {\vec{j}}}^l} c_{l,{\vec{j}}+\hat{\delta},\sigma}^\dagger c_{l,{\vec{j}},\sigma} + e^{-iA_{\hat{\delta}, {\vec{j}}}^l} c_{l,{\vec{j}},\sigma}^\dagger c_{l,{\vec{j}}+\hat{\delta},\sigma} \Big) \nonumber \\
       && -t\sum_{{\vec{j}}, \sigma}\Big(e^{iA_{z,{\vec{j}}}} c_{1,{\vec{j}},\sigma}^\dagger c_{2,{\vec{j}},\sigma} + e^{-iA_{z,{\vec{j}}}} c_{2,{\vec{j}},\sigma}^\dagger c_{1,{\vec{j}},\sigma} \Big) \nonumber \\
       && -\mu \sum_{{\vec{j}},l,\sigma}c_{l,{\vec{j}},\sigma}^\dagger c_{l,{\vec{j}},\sigma} 
       \\
       && \mathcal{H_I} =
        -U \sum_{{\vec{j}},l} c_{l,{\vec{j}},\uparrow}^\dagger c_{l,{\vec{j}},\downarrow}^\dagger c_{l,{\vec{j}},\downarrow} c_{l,{\vec{j}},\uparrow}. 
%    \end{split}
%\end{equation}
\end{eqnarray}
The layers are taken to be parallel to the x-y plane,
$l=\pm$ and $\sigma=\uparrow, \downarrow$ are the layer and spin indices, ${\vec{j}} \equiv (j_x,j_y)$ labels the lattice sites in a plane, and $\hat{\delta}=\hat{x}, \hat{y}$ denotes in-plane unit vectors. The operator $c_{l,{\vec{j}},\sigma}$ annihilates an electron with spin $\sigma$ at site ${\vec{j}}$ in layer $l$. The magnetic field is taken as $\vec{B} = B \hat{y}$. It is incorporated through $A_{\delta,{\vec{j}}}^l$ and $A_{z,{\vec{j}}}$, which are the Peierls phases for intra- and inter-layer hoppings,  determined by the condition that the phases around each plaquette add to $2\pi \Phi_p/\Phi_0$, where $\Phi_p$ is the flux per plaquette and $\Phi_0 = h/e$ is the flux quantum. 
We will work with the gauge choice, $A_{\hat{x},{\vec{j}}}^l = l\pi \times \Phi_p/\Phi_0$, and $A_{\hat{y},{\vec{j}}}^l = 0=A_{z,{\vec{j}}}$. Note that this gauge choice does not break the periodicity of the lattice, implying that the in-plane magnetic field does not break the translational symmetry; in other words, we do not need to define a magnetic unit cell (MUC) and the flux $\Phi$ can take arbitrary values.   As shown in the Supplementary Material, this model describes the low energy physics of bilayer Ising superconductors for which the spin is pinned to the out-of-plane directions and therefore the effect of Zeeman coupling is negligible. 

\begin{figure*}[t]
    \includegraphics[width = \textwidth]{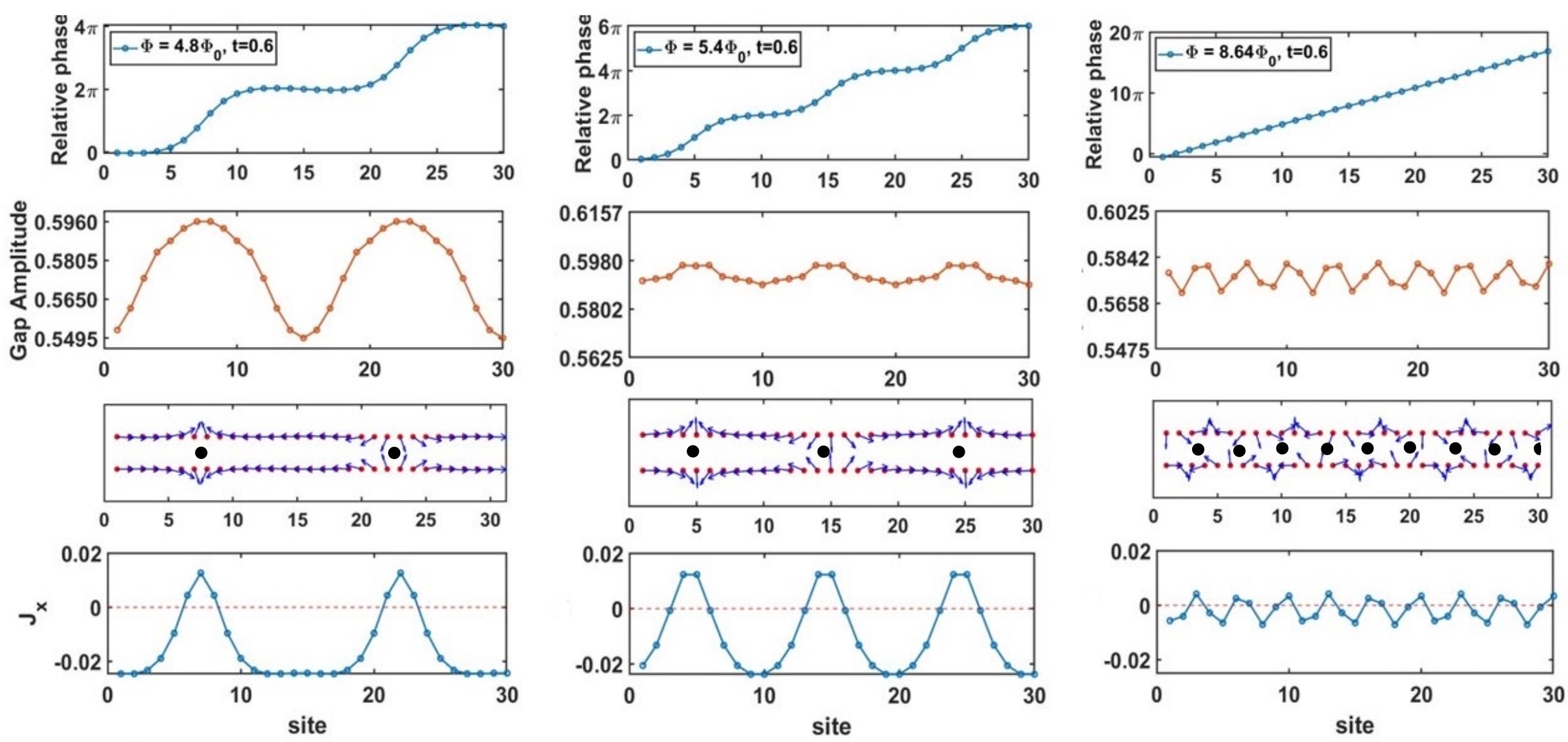}
    \caption{BCS stripe phase for a system of length $N=60$ for different values of the magnetic field. (For clarity, we show only half of the system; this behavior repeats in the other half.) For small $\Phi/\Phi_0$ the state (not shown) has zero relative phase, spatially uniform gap, equal and opposite currents in two layers, and no interlayer current. The first column corresponds to a magnetic flux $\Phi$ just above the critical value, the second to an intermediate $\Phi$, and the third to a high $\Phi$. Spatial variations are shown for: (a-c) the relative phase (difference between the phases in two layers); (d-f) the gap amplitude; (g-i) the phase of $\Delta$ [determined up to a global U(1) rotation]; and (j-l) the current density $J_x$ in the top layer. The magnitudes of the gap are same in both layers. The current in the bottom layer is $-J_x$, from which the interlayer current may be deduced. The black dots in (g-i) show the positions of interlayer vortices in the gap parameter (determined up to a global translation). The parameters chosen are $U=4, \mu=-3, T=0K, t=0.6$. Note that for a large $\Phi$ the relative phase varies linearly with the position and the gap amplitude becomes nearly constant.}
    \label{fig:fflo}
\end{figure*}

The mean field BdG Hamiltonian $H_{\rm MF}$, bilinear in electron operators, is obtained by replacing the $\mathcal{H_I}$ term by
%last term of Eq.(\ref{hamiltonian}) by
\begin{equation} \label{MF}
        \mathcal{H_I} \rightarrow 
%&-U\sum_{j,l}\Big(\langle c_{l,j,\downarrow} c_{l,j,\uparrow}\rangle c_{l,j,\uparrow}^\dagger c_{l,j,\downarrow}^\dagger + \\&\langle c_{l,j,\uparrow}^\dagger c_{l,j,\downarrow}^\dagger \rangle c_{l,j,\downarrow} c_{l,j,\uparrow} - \langle c_{l,j,\downarrow} c_{l,j,\uparrow}\rangle \langle c_{l,j,\uparrow}^\dagger c_{l,j,\downarrow}^\dagger \rangle \Big) \\ =& 
        -\sum_{l,{\vec{j}}}\Big(\Delta_{l,{\vec{j}}} c_{l,{\vec{j}},\uparrow}^\dagger c_{l,{\vec{j}},\downarrow}^\dagger + \Delta_{l,{\vec{j}}}^\star c_{l,{\vec{j}},\downarrow} c_{l,{\vec{j}},\uparrow} - \frac{|\Delta_{l,{\vec{j}}}|^2}{U} \Big)
\end{equation}
\begin{equation} \label{selfcon}
    \Delta_{l,{\vec{j}}} = U\langle c_{l,{\vec{j}},\downarrow} c_{l,{\vec{j}},\uparrow}\rangle.
\end{equation}
where $\langle \cdots \rangle$ denotes thermal average.
We solve for this Hamiltonian for a square system of size $N\times N$ in the x-y plane. We take a unit cell of size $1$ in the $\hat{y}$ direction, and let it span the entire system ($N$ sites) in the $\hat{x}$ direction. So, counting two layers, we have $N\times 1\times 2$ sites in the unit cell. We assume anti-periodic boundary conditions in both $\hat{x}$ and $\hat{y}$ direction. We perform our calculation for various $N$ values and extrapolate to $\frac{1}{N} \rightarrow 0$ to obtain the thermodynamic limits for various quantities (Fig.~\ref{fig:phasediagram} (b)).

To go to the momentum space, we write ${\vec{j}} = \vec{R}_{{\vec{j}}} + \alpha \hat{x}$, where $\vec{R}_{{\vec{j}}} = j_y \hat{y}$ is the position vector of the unit cell containing the site ${\vec{j}}$, and $\alpha=1\cdots N$ is the in-plane site index within the unit cell. We then define the transformation: $c_{l,{\vec{j}},\sigma} = \frac{1}{N}\sum_{\vec{k}} e^{i\vec{k}\cdot \vec{R}_{{\vec{j}}}} c_{l,\alpha, \sigma}(\vec{k})$, where $\vec{k} = \frac{\pi}{NL} n_y \hat{y}$; $n_y = -(N-1), -(N-3) \ldots  (N-1)$ ($n_y$ only takes odd integer values). Defining  an $8N$ dimensional vector $\Psi(\vec{k})=[\{c_{l,\alpha,\uparrow}(\vec{k})\}, \{c_{l,\alpha,\downarrow}(\vec{k})\}, \{c_{l,\alpha,\uparrow}^\dagger(-\vec{k})\}, \{c^{\dagger}_{l,\alpha,\downarrow}(-\vec{k})\}]^{\rm T}$, we obtain (closely following Ref.~\cite{Schirmer24})
\begin{eqnarray}
%\begin{equation} 
\label{hmf}
%    \begin{split}
        H_{\rm MF} &=& \frac{1}{2} \sum_{\vec{k}} \Psi^\dagger(\vec{k}) \hat{\mathcal{H}}_{\rm BdG}(\vec{k}) \Psi(\vec{k}) + \frac{1}{2}\sum_{\vec{k}} {\rm Tr}[H_0(\vec{k})] \nonumber\\ & + &\sum_{\vec{j},l}\frac{|\Delta_{l,\vec{j}}|^2}{U}, 
%    \end{split}
%\end{equation}
\end{eqnarray}
where $\hat{\mathcal{H}}_{\rm BdG}$ is an $8N\times 8N$ BdG matrix and $H_0(\vec{k})$ is the single particle part of the Hamiltonian in Fourier space. The eigen-energy spectrum of $H_{\rm MF}$ can be obtained by the diagonalization of $\hat{\mathcal{H}}_{\rm BdG}$.

To obtain self-consistent solutions we (i) begin with an initial guess $\Delta^{\rm guess}_{l,\alpha} = |\Delta| e^{\frac{iln \pi \alpha}{N}}$ on the site $\alpha=1,\cdots N$, which corresponds to $n$ vortices in the entire system; (ii) construct the BdG Hamiltonian to calculate $\Delta_{l,\alpha}$ using Eq.~(\ref{selfcon}); (iii) use this as our new guess; and (iv) repeat this process until we get a self consistent result, defined by the condition that the relative difference between the absolute values of the gaps in two successive steps is less than some tolerance (typically taken to be $10^{-5}$) at each site. We perform this for all values of $0\leq n\leq {\rm Int}[2\Phi/\Phi_0]+1$ and take the self-consistent solution with the lowest free energy ${\cal F}=\langle H_{\rm MF}-TS \rangle$, where the thermal average of the first term on the right hand side of Eq.~(\ref{hmf}) is given by $\frac{1}{2} \sum_{\vec{k}, \beta}E_{\beta}(\vec{k}) n_F(E_{\beta}(\vec{k}))$ with $n_F(E)=(e^{E/(k_BT)}+1)^{-1}$, and 
$\langle S \rangle = -k_B \sum_{k,\beta} [( n_F(E_{\beta}(\vec{k})) \log n_F(E_{\beta}(\vec{k}))$ $+(1-n_F(E_{\beta}(\vec{k}))) \log (1-n_F(E_{\beta}(\vec{k})))]$. 
We have tested the stability of the lowest energy solution by adding a random variation to the initial guess. Our solutions have zero net current, as required by Bloch's theorem~\cite{Watanabe2019}.

The in-plane current and that between two layers are given by $J_{\delta,l}(\vec{r}_{\vec{j}}) = -\frac{\partial \mathcal{H}}{\partial A_{\delta, l}}$ and $J_z(\vec{r}_{\vec{j}}) = -\frac{\partial \mathcal{H}}{\partial A_z}$, respectively. The average current densities in the $\hat{x}$ and $\hat{z}$ directions are
{\small
\begin{equation}
    \left\langle J_{x,\vec{j}}^l \right\rangle = i\sum_\sigma \left( e^{iA_{\hat{x}, \vec{j}}^l} \left\langle c_{l,\vec{j}+\hat{x},\sigma}^\dagger c_{l,\vec{j},\sigma}\right\rangle - e^{-iA_{\hat{x}, \vec{j}}^l} \left\langle c_{l,\vec{j},\sigma}^\dagger c_{l,\vec{j}+\hat{x},\sigma} \right\rangle \right) \nonumber
\end{equation}}

\begin{equation}
    \left\langle J_{z,\vec{j}} \right\rangle = it\sum_{\sigma}\left(e^{iA_{z,\vec{j}}} \left\langle c_{1,\vec{j},\sigma}^\dagger c_{2,\vec{j},\sigma}\right\rangle - e^{-iA_{z,\vec{j}}} \left\langle c_{2,\vec{j},\sigma}^\dagger c_{1,\vec{j},\sigma} \right\rangle \right) \nonumber
\end{equation}
and the average current density in the $\hat{y}$ direction vanishes.
The current density satisfies the continuity equation $\left\langle \nabla \cdot\vec{J} \right\rangle = \langle \partial_t \rho \rangle = i\langle[\mathcal{H},\rho] \rangle = 0$.

{\it Phase diagram - } We first consider the $T=0$ phase diagram and obtain for a given $N$ the lowest energy state to determine the critical values of $\Phi/\Phi_0$ below which the BCS phase survives. We obtain the thermodynamic limit by extrapolation. (The number of vortices in the ground state varies as a function of $N$.) The resulting phase diagram is shown in Fig.~\ref{fig:phasediagram}(c) as a function of $B$ and $t$. 

To bring out the nature of the ground state, an important quantity is the relative phase, namely the gauge-invariant difference between phases of the order parameter at two sites that lie directly across one another (i.e. have the same $x-y$ coordinates) in the two layers. The Hamiltonian is invariant under the gauge transformation: $c_{l,j,\sigma}\rightarrow \tilde{c}_{l,j,\sigma} =e^{i\phi_{j,l}}c_{l,j,\sigma}$, 
$\Delta_{j,l}\rightarrow\tilde{\Delta}_{j,l} = e^{2i\phi_{j,l}}\Delta_{j,l}$,
$A_{\hat{\delta}, {\vec{j}}}^l\rightarrow \tilde{A}_{\hat{\delta}, {\vec{j}}}^l = A_{\hat{\delta}, {\vec{j}}}^l - \phi_{\vec{j}+\hat{\delta},l} + \phi_{\vec{j},l}$, and $A_{z,j}\rightarrow \tilde{A}_{z,\vec{j}} = A_{z,j} - \phi_{\vec{j},+} + \phi_{\vec{j},-}$ (the last two are needed to ensure the invariance of the hopping term).  
The gauge invariant relative phase is defined as
${\rm arg}(\Delta_{+,\alpha}/\Delta_{-,\alpha}) + 2A_{z,\alpha}$ with $A_{z,\alpha}=0$ for our chosen gauge.

For sufficiently small $\Phi$, the ground state is in the ``BCS phase."  Here the relative phase is uniformly zero (the phases in the two layers are locked by interlayer tunneling and the gap can be taken to be real); the gap amplitude is constant; there are no interlayer supercurrents; and there is no spontaneous net currents, although the top and the bottom layers have equal and opposite diamagnetic currents\cite{Qiu22}.

The BCS phase survives up to a critical magnetic flux. Fig.~\ref{fig:fflo}(a) depicts the relative phase for a magnetic flux just above the critical value. Evidently, there are regions where the relative phase remains nearly constant separated by regions where it rapidly changes by $2\pi$. Thus, rather than a relative phase linearly varying with $x$, the system finds it energetically favorable to create BCS stripes running parallel to the direction of the magnetic field.  Associated with the formation of stripes is a periodic variation in the amplitude of the gap Fig.~\ref{fig:fflo}(d), and also the phase of the gap Fig.~\ref{fig:fflo}(g). The stripe phase also has a complex current pattern. Fig.~\ref{fig:fflo}(j) displays the current $j_x$ in the top layer; the current in the bottom layer is $-j_x$ and the interlayer current can be deduced from current conservation. The second column in Fig.~\ref{fig:fflo} depicts these quantities for an intermediate flux and the third for a large flux. In the limit of large flux, we find the relative phase changes linearly with $x$, recovering the layer helical phase in Ref.\cite{Liu17}, with a net change of about $4\pi \Phi/\Phi_0$ across the system corresponding to two vortices per flux quantum, and the gap amplitude is nearly constant. 

Intuitively, the BCS stripe phase can be viewed as arising from a competition between the inter-layer tunneling and the in-plane magnetic flux. While the inter-layer tunneling tends to lock the SC order parameter phases between two layers, thus favoring the BCS phase, the in-plane magnetic flux tends to drive a variation of this relative phase by introducing inter-layer vortices to form the layer helical phase. The highly non-uniform BCS stripe phase emerges as a compromise between these two competing tendencies. 

We have also evaluated the finite temperature phase diagram, shown in Fig.~\ref{fig:phasediagram}(d) for $t=0.5$. Fig.~\ref{fig:phasediagram}(e) shows how the gap amplitude varies as a function of T, where the gap  $|\Delta|$ in the stripe phase denotes the spatial average of the gap amplitude. The transitions from either the BCS or the stripe phase to the normal phase follow the standard behavior. The average gap amplitude varies smoothly across the transition from the BCS to the stripe phase because the number of vortices per flux quantum rises continuously from zero. 
Fig.~\ref{fig:phasediagram}(f) shows the behavior of the relative phase near the BCS to stripe phase boundary as a function of $T$; the phase becomes more linear as T approaches T$_{\rm c}$, consistent with Ref.\cite{Liu17}. 

Finally, a remark is in order on the gauge invariant pair momentum given by $\vec{p} = \vec{\nabla}\phi-2\vec{A}$, where $\Delta = |\Delta|e^{i\phi}$. In our lattice model this can be written as $p_{x,\vec{j}}^l = \phi_{l,\vec{j}+\hat{x}} - \phi_{l,\vec{j}} - 2A^l_{\hat{x},\vec{j}}$. Using this definition, we see that in the BCS phase, though there is no phase modulation of the gap parameter, there is an equal and opposite constant momentum in the top and bottom layers, which results in equal and opposite diamagnetic currents in the two layers. At high magnetic fields, there is equal and opposite contribution from both the phase and the gauge term in both layers. Since for high magnetic fields, $\phi = lq \cdot x$ with constant $q$ ($l=\pm$), the gauge invariant momentum is spatially uniform and equal to zero. This is not the case, however, in the stripe phase (Fig.~\ref{fig:fflo} (g) and (h)).

\textit{Discussion -}
The interplay between superconductivity and magnetic field in 2D systems has been of interest in numerous contexts, including the possibility of topological superconductivity supporting Majorana zero modes~\cite{Stone2006,Oshikawa2007,Nayak2008,Sau2010a,Stanescu2010,Alicea2012,Hung2013,Beenakker2013,Zhou2013,Biswas2013,Sarma2015,Liu2015,Murray2015,Smith2016}. Here we predict a stripe phase in a bilayer superconductor exposed to a parallel magnetic field.

An earlier Landau-Ginzburg treatment of this problem by Liu~\cite{Liu17} assumes the layer helical phase ansatz $\Delta_l(\vec{r})=|\Delta|e^{ilqx}$ where $l=\pm$ is the layer index and $q$ is spatially uniform. He finds $q_c = \frac{2eBz_0}{2\hbar}$ in the limit of large $B$, where $z_0$ is the distance between two layers; the relative phase difference $2q_cx$ changes by $2\pi$ in a distance $\Delta x=2\pi / 2q_c$, which contains $B z_0\Delta x =\Phi_0/2$ flux quanta, thus producing twice as many vortices as the number of flux quanta. In contrast, the highly non-uniform stripe phase found in this work cannot be represented by layer helical phase with a single $q$ in each layer. Qui and Zhou~\cite{Qiu22} have also studied this problem within the Landau-Ginzburg formalism, while assuming that the gap function has the periodicity of an MUC containing one flux quantum (the ratio of the number of vortices to the number of flux quanta can take only discrete values in this model); this periodicity results in a Bloch wave solution for the superconducting order parameter and a commensurate-incommensurate phase transition. The BCS stripe phase in our work does {\it not} necessarily belong to the Bloch wave solution as it in general does not respect the periodicity of the MUC.

By obtaining the self-consistent solutions of the BdG equations, we have shown that the application of a parallel $B$ causes a transition into a stripe phase. We thus predict that the phase transition observed in 2H-stacked NbSe$_2$ in Ref.~\cite{Wan2023} is into a highly nonuniform stripe phase, which adiabatically evolves into the inter the helical phase of Ref.\cite{Liu17} for large $B$. This stripe phase may be identified most directly by measuring the oscillations in the amplitude of the gap, which are predicted to be strongest at low temperatures near the phase boundary.

{\it Acknowledgments -} UN, JS (at Penn State) and JKJ were supported in part by the U. S. Department of Energy, Office of Basic Energy Sciences, under Grant no. DE-SC-0005042. Work by JS and ER was supported by the U.S. Department of Energy (DOE), Office of Science, Basic Energy Sciences (BES) under Award DE-SC0022245. C.-X.L 
acknowledges the support from the NSF through The Pennsylvania State University Materials Research Science
and Engineering Center [DMR-2011839]. We acknowledge Advanced CyberInfrastructure computational resources provided by The Institute for CyberScience at The Pennsylvania State University.

\begin{appendix}

\section{Appendix: Mapping bilayer TMD system with Ising spin-orbit coupling to our model}
In this section, we will demonstrate that the model we used can serve as the low-energy theory of the bilayer transition-metal-dichalcogenide (TMD) system with Ising spin-orbit coupling. We start from the continuous model for a TMD monolayer. The single particle Hamiltonian for the layer $l$ ($l = 1,2$) at zero magnetic field reads
\begin{equation}
    \mathcal{H}_{0,l}(\vec{p} = \epsilon \vec{K} + \vec{k}) = \xi_k + (-1)^l\epsilon\beta_{SOC}s_z,
\end{equation}
where $\xi_k$ is the kinetic energy term, $\epsilon=\pm$ is the valley index, $\beta_{SOC}$ is the Ising spin-orbit coupling strength, and $s_z$ is the Pauli matrix for the spin degree of freedom. In the second quantization notation with the electron annihilation operator denoted as $c_{\vec{k},\epsilon,l,s}$ ($s$ being the spin index), the Hamiltonian takes the form
\begin{equation}
    \hat{H}_0 = \sum_{\vec{k},\epsilon,s,s^\prime} c^\dagger_{\vec{k},\epsilon,l,s} [\mathcal{H}_{0,l}(\epsilon\vec{K} + \vec{k})]_{s,s^\prime} c_{\vec{k},\epsilon,l,s^\prime}.
\end{equation}
One can see that this Hamiltonian has a diagonal form with the spin-valley-layer locking, namely $\expval{\hat{H}_{0,l}}{+\uparrow} = \xi_k + (-1)^l\beta_{SOC}$, $\expval{\hat{H}_{0,l}}{+\downarrow} = \xi_k - (-1)^l\beta_{SOC}$, $\expval{\hat{H}_{0,l}}{-\uparrow} = \xi_k - (-1)^l\beta_{SOC}$, and $\expval{\hat{H}_{0,l}}{+\uparrow} = \xi_k + (-1)^l\beta_{SOC}$ for non-zero Hamiltonian matrix elements. 
Since we only wish to consider the lowest energy bands and assume $\beta_{SOC}>0$, we project the full Hamiltonian onto the basis $A = \{ \ket{l=1,\epsilon=+, \uparrow}$, $\ket{l=1,\epsilon=-,\downarrow}$, $\ket{l=2,\epsilon=+,\downarrow}$, $\ket{l=2,\epsilon=-,\uparrow} \}$. The low energy Hamiltonian then becomes:
$$\hat{H}^{low}_0 = \sum_{\vec{k},\alpha \in A}(\xi_k - \beta_{SOC})c^\dagger_{\vec{k},\alpha}c_{\vec{k},\alpha}.$$
Note that the spin-valley indices are opposite for the two layers in the above basis wavefunctions, and this will have substantial influence on the inter-layer tunneling.

The inter-layer tunneling term in the original basis (before projecting onto the lowest energy bands) preserves both spin and valley:
$$\hat{H}_t = -t\sum_{\vec{k},\epsilon,s} \Big( c^\dagger_{\vec{k},\epsilon,1,s} c_{\vec{k},\epsilon,2,s} + c^\dagger_{\vec{k},\epsilon,2,s} c_{\vec{k},\epsilon,1,s} \Big).$$
But, for $l=1$ and $l=2$, the spin and valley indices are opposite for low energy bands in the basis set $A$. Therefore the direct tunneling from this term in the low energy theory is zero. Inter-layer tunneling between low-energy bands can, however, be mediated by the combination of $\hat{H}_t$ and the Zeeman coupling from the in-plane magnetic field.

We consider the in-plane magnetic field along the $\hat{x}$ direction. The Zeeman coupling is then given by 
$$\hat{H}_Z = gB_x\sum_{\vec{k},l,\epsilon,s} c^\dagger_{\vec{k},\epsilon,l,s} (\sigma_x)_{s,s^\prime} c_{\vec{k},\epsilon,l,s^\prime},$$
where $g$ is the g-factor. 
From the Hamiltonian forms of $\hat{H}_t$ and $\hat{H}_Z$, we see that while the inter-layer tunneling term preserves the spin-valley index while changing the layer index, the Zeeman term for an in-plane magnetic field changes the spin index while preserving the other two. Thus, we get an effective tunneling between the lowest energy bands from the second order term in perturbation theory:
$$\hat{H}_t^{low} = \frac{tgB_x}{\beta_{SOC}}\sum_{\vec{k}}\Big[ c^\dagger_{\vec{k},+,1,\uparrow} c_{\vec{k},+,2,\downarrow} +  c^\dagger_{\vec{k},-,1,\downarrow} c_{\vec{k},-,2,\uparrow} + h.c. \Big].$$
The effective low energy theory is given by:
$$\hat{H}^{low} = \hat{H}^{low}_0 + \hat{H}^{low}_t. $$

Due to the redundancy in the labelling of the basis wavefunctions, we can remove the valley index in the notation, so that  
$\ket{l=1,\epsilon=+,\uparrow} \rightarrow \ket{l=1,\uparrow}$, $\ket{l=1,\epsilon=-,\downarrow} \rightarrow \ket{l=1,\downarrow}$, $\ket{l=2,\epsilon=+,\downarrow} \rightarrow \ket{l=2,\uparrow}$, $\ket{l=2,\epsilon=-,\uparrow} \rightarrow \ket{l=2,\downarrow}$, and $\frac{tgB_x}{\beta_{SOC}} \rightarrow t$. The resulting Hamiltonian is a model for the spinful electrons in a bilayer system. We may include the orbital effect of magnetic field by performing the Peierls substitution and then implement the lattice regularization for $\hat{H}^{low}$ to transform it to the lattice model, and with that, we can recover the noninteracting Hamiltonian $\mathcal{H}_0$ in the main text. Thus, the in-plane Zeeman effect only provides a correction to the inter-layer tunneling parameter $t$.

\end{appendix}

\bibliography{Schirmer.bib}

\begin{thebibliography}{34}%
\makeatletter
\providecommand \@ifxundefined [1]{%
 \@ifx{#1\undefined}
}%
\providecommand \@ifnum [1]{%
 \ifnum #1\expandafter \@firstoftwo
 \else \expandafter \@secondoftwo
 \fi
}%
\providecommand \@ifx [1]{%
 \ifx #1\expandafter \@firstoftwo
 \else \expandafter \@secondoftwo
 \fi
}%
\providecommand \natexlab [1]{#1}%
\providecommand \enquote  [1]{``#1''}%
\providecommand \bibnamefont  [1]{#1}%
\providecommand \bibfnamefont [1]{#1}%
\providecommand \citenamefont [1]{#1}%
\providecommand \href@noop [0]{\@secondoftwo}%
\providecommand \href [0]{\begingroup \@sanitize@url \@href}%
\providecommand \@href[1]{\@@startlink{#1}\@@href}%
\providecommand \@@href[1]{\endgroup#1\@@endlink}%
\providecommand \@sanitize@url [0]{\catcode `\\12\catcode `\$12\catcode `\&12\catcode `\#12\catcode `\^12\catcode `\_12\catcode `\%12\relax}%
\providecommand \@@startlink[1]{}%
\providecommand \@@endlink[0]{}%
\providecommand \url  [0]{\begingroup\@sanitize@url \@url }%
\providecommand \@url [1]{\endgroup\@href {#1}{\urlprefix }}%
\providecommand \urlprefix  [0]{URL }%
\providecommand \Eprint [0]{\href }%
\providecommand \doibase [0]{http://dx.doi.org/}%
\providecommand \selectlanguage [0]{\@gobble}%
\providecommand \bibinfo  [0]{\@secondoftwo}%
\providecommand \bibfield  [0]{\@secondoftwo}%
\providecommand \translation [1]{[#1]}%
\providecommand \BibitemOpen [0]{}%
\providecommand \bibitemStop [0]{}%
\providecommand \bibitemNoStop [0]{.\EOS\space}%
\providecommand \EOS [0]{\spacefactor3000\relax}%
\providecommand \BibitemShut  [1]{\csname bibitem#1\endcsname}%
\let\auto@bib@innerbib\@empty
%</preamble>
\bibitem [{\citenamefont {Lilly}\ \emph {et~al.}(1999)\citenamefont {Lilly}, \citenamefont {Cooper}, \citenamefont {Eisenstein}, \citenamefont {Pfeiffer},\ and\ \citenamefont {West}}]{Lilly99}%
  \BibitemOpen
  \bibfield  {author} {\bibinfo {author} {\bibfnamefont {M.~P.}\ \bibnamefont {Lilly}}, \bibinfo {author} {\bibfnamefont {K.~B.}\ \bibnamefont {Cooper}}, \bibinfo {author} {\bibfnamefont {J.~P.}\ \bibnamefont {Eisenstein}}, \bibinfo {author} {\bibfnamefont {L.~N.}\ \bibnamefont {Pfeiffer}}, \ and\ \bibinfo {author} {\bibfnamefont {K.~W.}\ \bibnamefont {West}},\ }\href {\doibase 10.1103/PhysRevLett.82.394} {\bibfield  {journal} {\bibinfo  {journal} {Physical Review Letters}\ }\textbf {\bibinfo {volume} {82}},\ \bibinfo {pages} {394} (\bibinfo {year} {1999})}\BibitemShut {NoStop}%
\bibitem [{\citenamefont {Fogler}\ \emph {et~al.}(1996)\citenamefont {Fogler}, \citenamefont {Koulakov},\ and\ \citenamefont {Shklovskii}}]{Fogler96}%
  \BibitemOpen
  \bibfield  {author} {\bibinfo {author} {\bibfnamefont {M.~M.}\ \bibnamefont {Fogler}}, \bibinfo {author} {\bibfnamefont {A.~A.}\ \bibnamefont {Koulakov}}, \ and\ \bibinfo {author} {\bibfnamefont {B.~I.}\ \bibnamefont {Shklovskii}},\ }\href {\doibase 10.1103/PhysRevB.54.1853} {\bibfield  {journal} {\bibinfo  {journal} {Physical Review B}\ }\textbf {\bibinfo {volume} {54}},\ \bibinfo {pages} {1853} (\bibinfo {year} {1996})}\BibitemShut {NoStop}%
\bibitem [{\citenamefont {Emery}\ \emph {et~al.}(1999)\citenamefont {Emery}, \citenamefont {Kivelson},\ and\ \citenamefont {Tranquada}}]{Emery99}%
  \BibitemOpen
  \bibfield  {author} {\bibinfo {author} {\bibfnamefont {V.~J.}\ \bibnamefont {Emery}}, \bibinfo {author} {\bibfnamefont {S.~A.}\ \bibnamefont {Kivelson}}, \ and\ \bibinfo {author} {\bibfnamefont {J.~M.}\ \bibnamefont {Tranquada}},\ }\href {\doibase 10.1073/pnas.96.16.8814} {\bibfield  {journal} {\bibinfo  {journal} {Proceedings of the National Academy of Sciences}\ }\textbf {\bibinfo {volume} {96}},\ \bibinfo {pages} {8814} (\bibinfo {year} {1999})}\BibitemShut {NoStop}%
\bibitem [{\citenamefont {Liu}(2017)}]{Liu17}%
  \BibitemOpen
  \bibfield  {author} {\bibinfo {author} {\bibfnamefont {C.-X.}\ \bibnamefont {Liu}},\ }\href {\doibase 10.1103/PhysRevLett.118.087001} {\bibfield  {journal} {\bibinfo  {journal} {Physical Review Letters}\ }\textbf {\bibinfo {volume} {118}},\ \bibinfo {pages} {087001} (\bibinfo {year} {2017})}\BibitemShut {NoStop}%
\bibitem [{\citenamefont {Fulde}\ and\ \citenamefont {Ferrell}(1964)}]{Fulde64}%
  \BibitemOpen
  \bibfield  {author} {\bibinfo {author} {\bibfnamefont {P.}~\bibnamefont {Fulde}}\ and\ \bibinfo {author} {\bibfnamefont {R.~A.}\ \bibnamefont {Ferrell}},\ }\href {\doibase 10.1103/PhysRev.135.A550} {\bibfield  {journal} {\bibinfo  {journal} {Physical Review}\ }\textbf {\bibinfo {volume} {135}},\ \bibinfo {pages} {A550} (\bibinfo {year} {1964})}\BibitemShut {NoStop}%
\bibitem [{\citenamefont {Larkin}\ and\ \citenamefont {Ovchinnikov}(1965)}]{larkin1965nonuniform}%
  \BibitemOpen
  \bibfield  {author} {\bibinfo {author} {\bibfnamefont {A.}~\bibnamefont {Larkin}}\ and\ \bibinfo {author} {\bibfnamefont {Y.~N.}\ \bibnamefont {Ovchinnikov}},\ }\href@noop {} {\bibfield  {journal} {\bibinfo  {journal} {Soviet Physics-JETP}\ }\textbf {\bibinfo {volume} {20}},\ \bibinfo {pages} {762} (\bibinfo {year} {1965})}\BibitemShut {NoStop}%
\bibitem [{\citenamefont {Agterberg}(2012)}]{agterberg2012magnetoelectric}%
  \BibitemOpen
  \bibfield  {author} {\bibinfo {author} {\bibfnamefont {D.}~\bibnamefont {Agterberg}},\ }\href@noop {} {\bibfield  {journal} {\bibinfo  {journal} {Non-Centrosymmetric Superconductors: Introduction and Overview}\ ,\ \bibinfo {pages} {155}} (\bibinfo {year} {2012})}\BibitemShut {NoStop}%
\bibitem [{\citenamefont {Kinnunen}\ \emph {et~al.}(2018{\natexlab{a}})\citenamefont {Kinnunen}, \citenamefont {Baarsma}, \citenamefont {Martikainen},\ and\ \citenamefont {T{\"o}rm{\"a}}}]{kinnunen2018fulde}%
  \BibitemOpen
  \bibfield  {author} {\bibinfo {author} {\bibfnamefont {J.~J.}\ \bibnamefont {Kinnunen}}, \bibinfo {author} {\bibfnamefont {J.~E.}\ \bibnamefont {Baarsma}}, \bibinfo {author} {\bibfnamefont {J.-P.}\ \bibnamefont {Martikainen}}, \ and\ \bibinfo {author} {\bibfnamefont {P.}~\bibnamefont {T{\"o}rm{\"a}}},\ }\href@noop {} {\bibfield  {journal} {\bibinfo  {journal} {Reports on Progress in Physics}\ }\textbf {\bibinfo {volume} {81}},\ \bibinfo {pages} {046401} (\bibinfo {year} {2018}{\natexlab{a}})}\BibitemShut {NoStop}%
\bibitem [{\citenamefont {Bianchi}\ \emph {et~al.}(2003)\citenamefont {Bianchi}, \citenamefont {Movshovich}, \citenamefont {Capan}, \citenamefont {Pagliuso},\ and\ \citenamefont {Sarrao}}]{Bianchi2003}%
  \BibitemOpen
  \bibfield  {author} {\bibinfo {author} {\bibfnamefont {A.}~\bibnamefont {Bianchi}}, \bibinfo {author} {\bibfnamefont {R.}~\bibnamefont {Movshovich}}, \bibinfo {author} {\bibfnamefont {C.}~\bibnamefont {Capan}}, \bibinfo {author} {\bibfnamefont {P.~G.}\ \bibnamefont {Pagliuso}}, \ and\ \bibinfo {author} {\bibfnamefont {J.~L.}\ \bibnamefont {Sarrao}},\ }\href {\doibase 10.1103/PhysRevLett.91.187004} {\bibfield  {journal} {\bibinfo  {journal} {Physical Review Letters}\ }\textbf {\bibinfo {volume} {91}},\ \bibinfo {pages} {187004} (\bibinfo {year} {2003})}\BibitemShut {NoStop}%
\bibitem [{\citenamefont {Matsuda}\ and\ \citenamefont {Shimahara}(2007)}]{Matsuda2007}%
  \BibitemOpen
  \bibfield  {author} {\bibinfo {author} {\bibfnamefont {Y.}~\bibnamefont {Matsuda}}\ and\ \bibinfo {author} {\bibfnamefont {H.}~\bibnamefont {Shimahara}},\ }\href {\doibase 10.1143/JPSJ.76.051005} {\bibfield  {journal} {\bibinfo  {journal} {Journal of the Physical Society of Japan}\ }\textbf {\bibinfo {volume} {76}},\ \bibinfo {pages} {051005} (\bibinfo {year} {2007})}\BibitemShut {NoStop}%
\bibitem [{\citenamefont {Partridge}\ \emph {et~al.}(2006)\citenamefont {Partridge}, \citenamefont {Li}, \citenamefont {Kamar}, \citenamefont {an~Liao},\ and\ \citenamefont {Hulet}}]{Partridge2006}%
  \BibitemOpen
  \bibfield  {author} {\bibinfo {author} {\bibfnamefont {G.~B.}\ \bibnamefont {Partridge}}, \bibinfo {author} {\bibfnamefont {W.}~\bibnamefont {Li}}, \bibinfo {author} {\bibfnamefont {R.~I.}\ \bibnamefont {Kamar}}, \bibinfo {author} {\bibfnamefont {Y.}~\bibnamefont {an~Liao}}, \ and\ \bibinfo {author} {\bibfnamefont {R.~G.}\ \bibnamefont {Hulet}},\ }\href {\doibase 10.1126/science.1122876} {\bibfield  {journal} {\bibinfo  {journal} {Science}\ }\textbf {\bibinfo {volume} {311}},\ \bibinfo {pages} {503} (\bibinfo {year} {2006})}\BibitemShut {NoStop}%
\bibitem [{\citenamefont {Parish}\ \emph {et~al.}(2007)\citenamefont {Parish}, \citenamefont {Baur}, \citenamefont {Mueller},\ and\ \citenamefont {Huse}}]{Parish2007}%
  \BibitemOpen
  \bibfield  {author} {\bibinfo {author} {\bibfnamefont {M.~M.}\ \bibnamefont {Parish}}, \bibinfo {author} {\bibfnamefont {S.~K.}\ \bibnamefont {Baur}}, \bibinfo {author} {\bibfnamefont {E.~J.}\ \bibnamefont {Mueller}}, \ and\ \bibinfo {author} {\bibfnamefont {D.~A.}\ \bibnamefont {Huse}},\ }\href {\doibase 10.1103/PhysRevLett.99.250403} {\bibfield  {journal} {\bibinfo  {journal} {Physical Review Letters}\ }\textbf {\bibinfo {volume} {99}},\ \bibinfo {pages} {250403} (\bibinfo {year} {2007})}\BibitemShut {NoStop}%
\bibitem [{\citenamefont {Kinnunen}\ \emph {et~al.}(2018{\natexlab{b}})\citenamefont {Kinnunen}, \citenamefont {Baarsma}, \citenamefont {Martikainen},\ and\ \citenamefont {Törmä}}]{Kinnunen2018}%
  \BibitemOpen
  \bibfield  {author} {\bibinfo {author} {\bibfnamefont {J.~J.}\ \bibnamefont {Kinnunen}}, \bibinfo {author} {\bibfnamefont {J.~E.}\ \bibnamefont {Baarsma}}, \bibinfo {author} {\bibfnamefont {J.-P.}\ \bibnamefont {Martikainen}}, \ and\ \bibinfo {author} {\bibfnamefont {P.}~\bibnamefont {Törmä}},\ }\href {\doibase 10.1088/1361-6633/aaa4ad} {\bibfield  {journal} {\bibinfo  {journal} {Reports on Progress in Physics}\ }\textbf {\bibinfo {volume} {81}},\ \bibinfo {pages} {046401} (\bibinfo {year} {2018}{\natexlab{b}})}\BibitemShut {NoStop}%
\bibitem [{\citenamefont {Zhao}\ \emph {et~al.}(2023)\citenamefont {Zhao}, \citenamefont {Debbeler}, \citenamefont {Kühne}, \citenamefont {Fecher}, \citenamefont {Gross},\ and\ \citenamefont {Smet}}]{Zhao2023centrosym}%
  \BibitemOpen
  \bibfield  {author} {\bibinfo {author} {\bibfnamefont {D.}~\bibnamefont {Zhao}}, \bibinfo {author} {\bibfnamefont {L.}~\bibnamefont {Debbeler}}, \bibinfo {author} {\bibfnamefont {M.}~\bibnamefont {Kühne}}, \bibinfo {author} {\bibfnamefont {S.}~\bibnamefont {Fecher}}, \bibinfo {author} {\bibfnamefont {N.}~\bibnamefont {Gross}}, \ and\ \bibinfo {author} {\bibfnamefont {J.}~\bibnamefont {Smet}},\ }\href {\doibase 10.1038/s41567-023-02202-4} {\bibfield  {journal} {\bibinfo  {journal} {Nature Physics}\ }\textbf {\bibinfo {volume} {19}},\ \bibinfo {pages} {1599} (\bibinfo {year} {2023})}\BibitemShut {NoStop}%
\bibitem [{\citenamefont {Wan}\ \emph {et~al.}(2023{\natexlab{a}})\citenamefont {Wan}, \citenamefont {Zheliuk}, \citenamefont {Yuan}, \citenamefont {Peng}, \citenamefont {Zhang}, \citenamefont {Liang}, \citenamefont {Zeitler}, \citenamefont {Wiedmann}, \citenamefont {Hussey}, \citenamefont {Palstra},\ and\ \citenamefont {Ye}}]{Wan23}%
  \BibitemOpen
  \bibfield  {author} {\bibinfo {author} {\bibfnamefont {P.}~\bibnamefont {Wan}}, \bibinfo {author} {\bibfnamefont {O.}~\bibnamefont {Zheliuk}}, \bibinfo {author} {\bibfnamefont {N.~F.~Q.}\ \bibnamefont {Yuan}}, \bibinfo {author} {\bibfnamefont {X.}~\bibnamefont {Peng}}, \bibinfo {author} {\bibfnamefont {L.}~\bibnamefont {Zhang}}, \bibinfo {author} {\bibfnamefont {M.}~\bibnamefont {Liang}}, \bibinfo {author} {\bibfnamefont {U.}~\bibnamefont {Zeitler}}, \bibinfo {author} {\bibfnamefont {S.}~\bibnamefont {Wiedmann}}, \bibinfo {author} {\bibfnamefont {N.~E.}\ \bibnamefont {Hussey}}, \bibinfo {author} {\bibfnamefont {T.~T.~M.}\ \bibnamefont {Palstra}}, \ and\ \bibinfo {author} {\bibfnamefont {J.}~\bibnamefont {Ye}},\ }\href {\doibase 10.1038/s41586-023-05967-z} {\bibfield  {journal} {\bibinfo  {journal} {Nature}\ }\textbf {\bibinfo {volume} {619}},\ \bibinfo {pages} {46} (\bibinfo {year} {2023}{\natexlab{a}})}\BibitemShut {NoStop}%
\bibitem [{\citenamefont {Croitoru}\ and\ \citenamefont {Buzdin}(2017)}]{Croitoru17}%
  \BibitemOpen
  \bibfield  {author} {\bibinfo {author} {\bibfnamefont {M.}~\bibnamefont {Croitoru}}\ and\ \bibinfo {author} {\bibfnamefont {A.}~\bibnamefont {Buzdin}},\ }\href {\doibase 10.3390/condmat2030030} {\bibfield  {journal} {\bibinfo  {journal} {Condensed Matter}\ }\textbf {\bibinfo {volume} {2}},\ \bibinfo {pages} {30} (\bibinfo {year} {2017})}\BibitemShut {NoStop}%
\bibitem [{\citenamefont {Schirmer}\ \emph {et~al.}(2024)\citenamefont {Schirmer}, \citenamefont {Jain},\ and\ \citenamefont {Liu}}]{Schirmer24}%
  \BibitemOpen
  \bibfield  {author} {\bibinfo {author} {\bibfnamefont {J.}~\bibnamefont {Schirmer}}, \bibinfo {author} {\bibfnamefont {J.~K.}\ \bibnamefont {Jain}}, \ and\ \bibinfo {author} {\bibfnamefont {C.~X.}\ \bibnamefont {Liu}},\ }\href {\doibase 10.1103/PhysRevB.109.134518} {\bibfield  {journal} {\bibinfo  {journal} {Physical Review B}\ }\textbf {\bibinfo {volume} {109}},\ \bibinfo {pages} {134518} (\bibinfo {year} {2024})}\BibitemShut {NoStop}%
\bibitem [{\citenamefont {Watanabe}(2019)}]{Watanabe2019}%
  \BibitemOpen
  \bibfield  {author} {\bibinfo {author} {\bibfnamefont {H.}~\bibnamefont {Watanabe}},\ }\href@noop {} {\bibfield  {journal} {\bibinfo  {journal} {Journal of Statistical Physics}\ }\textbf {\bibinfo {volume} {177}},\ \bibinfo {pages} {717} (\bibinfo {year} {2019})}\BibitemShut {NoStop}%
\bibitem [{\citenamefont {Qiu}\ and\ \citenamefont {Zhou}(2022)}]{Qiu22}%
  \BibitemOpen
  \bibfield  {author} {\bibinfo {author} {\bibfnamefont {G.-W.}\ \bibnamefont {Qiu}}\ and\ \bibinfo {author} {\bibfnamefont {Y.}~\bibnamefont {Zhou}},\ }\href {\doibase 10.1103/PhysRevB.105.L100506} {\bibfield  {journal} {\bibinfo  {journal} {Physical Review B}\ }\textbf {\bibinfo {volume} {105}},\ \bibinfo {pages} {L100506} (\bibinfo {year} {2022})}\BibitemShut {NoStop}%
\bibitem [{\citenamefont {Stone}\ and\ \citenamefont {Chung}(2006)}]{Stone2006}%
  \BibitemOpen
  \bibfield  {author} {\bibinfo {author} {\bibfnamefont {M.}~\bibnamefont {Stone}}\ and\ \bibinfo {author} {\bibfnamefont {S.-B.}\ \bibnamefont {Chung}},\ }\href@noop {} {\bibfield  {journal} {\bibinfo  {journal} {Physical Review B}\ }\textbf {\bibinfo {volume} {73}},\ \bibinfo {pages} {014505} (\bibinfo {year} {2006})}\BibitemShut {NoStop}%
\bibitem [{\citenamefont {Oshikawa}\ \emph {et~al.}(2007)\citenamefont {Oshikawa}, \citenamefont {Kim}, \citenamefont {Shtengel}, \citenamefont {Nayak},\ and\ \citenamefont {Tewari}}]{Oshikawa2007}%
  \BibitemOpen
  \bibfield  {author} {\bibinfo {author} {\bibfnamefont {M.}~\bibnamefont {Oshikawa}}, \bibinfo {author} {\bibfnamefont {Y.~B.}\ \bibnamefont {Kim}}, \bibinfo {author} {\bibfnamefont {K.}~\bibnamefont {Shtengel}}, \bibinfo {author} {\bibfnamefont {C.}~\bibnamefont {Nayak}}, \ and\ \bibinfo {author} {\bibfnamefont {S.}~\bibnamefont {Tewari}},\ }\href {\doibase https://doi.org/10.1016/j.aop.2006.08.001} {\bibfield  {journal} {\bibinfo  {journal} {Annals of Physics}\ }\textbf {\bibinfo {volume} {322}},\ \bibinfo {pages} {1477 } (\bibinfo {year} {2007})}\BibitemShut {NoStop}%
\bibitem [{\citenamefont {Nayak}\ \emph {et~al.}(2008)\citenamefont {Nayak}, \citenamefont {Simon}, \citenamefont {Stern}, \citenamefont {Freedman},\ and\ \citenamefont {Das~Sarma}}]{Nayak2008}%
  \BibitemOpen
  \bibfield  {author} {\bibinfo {author} {\bibfnamefont {C.}~\bibnamefont {Nayak}}, \bibinfo {author} {\bibfnamefont {S.~H.}\ \bibnamefont {Simon}}, \bibinfo {author} {\bibfnamefont {A.}~\bibnamefont {Stern}}, \bibinfo {author} {\bibfnamefont {M.}~\bibnamefont {Freedman}}, \ and\ \bibinfo {author} {\bibfnamefont {S.}~\bibnamefont {Das~Sarma}},\ }\href {\doibase 10.1103/RevModPhys.80.1083} {\bibfield  {journal} {\bibinfo  {journal} {Reviews of Modern Physics}\ }\textbf {\bibinfo {volume} {80}},\ \bibinfo {pages} {1083} (\bibinfo {year} {2008})}\BibitemShut {NoStop}%
\bibitem [{\citenamefont {Sau}\ \emph {et~al.}(2010)\citenamefont {Sau}, \citenamefont {Lutchyn}, \citenamefont {Tewari},\ and\ \citenamefont {Sarma}}]{Sau2010a}%
  \BibitemOpen
  \bibfield  {author} {\bibinfo {author} {\bibfnamefont {J.~D.}\ \bibnamefont {Sau}}, \bibinfo {author} {\bibfnamefont {R.~M.}\ \bibnamefont {Lutchyn}}, \bibinfo {author} {\bibfnamefont {S.}~\bibnamefont {Tewari}}, \ and\ \bibinfo {author} {\bibfnamefont {S.~D.}\ \bibnamefont {Sarma}},\ }\href@noop {} {\bibfield  {journal} {\bibinfo  {journal} {Physical Review Letters}\ }\textbf {\bibinfo {volume} {104}},\ \bibinfo {pages} {040502} (\bibinfo {year} {2010})}\BibitemShut {NoStop}%
\bibitem [{\citenamefont {Stanescu}\ \emph {et~al.}(2010)\citenamefont {Stanescu}, \citenamefont {Sau}, \citenamefont {Lutchyn},\ and\ \citenamefont {Sarma}}]{Stanescu2010}%
  \BibitemOpen
  \bibfield  {author} {\bibinfo {author} {\bibfnamefont {T.~D.}\ \bibnamefont {Stanescu}}, \bibinfo {author} {\bibfnamefont {J.~D.}\ \bibnamefont {Sau}}, \bibinfo {author} {\bibfnamefont {R.~M.}\ \bibnamefont {Lutchyn}}, \ and\ \bibinfo {author} {\bibfnamefont {S.~D.}\ \bibnamefont {Sarma}},\ }\href@noop {} {\bibfield  {journal} {\bibinfo  {journal} {Physical Review B}\ }\textbf {\bibinfo {volume} {81}},\ \bibinfo {pages} {241310} (\bibinfo {year} {2010})}\BibitemShut {NoStop}%
\bibitem [{\citenamefont {Alicea}(2012)}]{Alicea2012}%
  \BibitemOpen
  \bibfield  {author} {\bibinfo {author} {\bibfnamefont {J.}~\bibnamefont {Alicea}},\ }\href@noop {} {\bibfield  {journal} {\bibinfo  {journal} {Reports on Progress in Physics}\ }\textbf {\bibinfo {volume} {75}},\ \bibinfo {pages} {076501} (\bibinfo {year} {2012})}\BibitemShut {NoStop}%
\bibitem [{\citenamefont {Hung}\ \emph {et~al.}(2013)\citenamefont {Hung}, \citenamefont {Ghaemi}, \citenamefont {Hughes},\ and\ \citenamefont {Gilbert}}]{Hung2013}%
  \BibitemOpen
  \bibfield  {author} {\bibinfo {author} {\bibfnamefont {H.-H.}\ \bibnamefont {Hung}}, \bibinfo {author} {\bibfnamefont {P.}~\bibnamefont {Ghaemi}}, \bibinfo {author} {\bibfnamefont {T.~L.}\ \bibnamefont {Hughes}}, \ and\ \bibinfo {author} {\bibfnamefont {M.~J.}\ \bibnamefont {Gilbert}},\ }\href@noop {} {\bibfield  {journal} {\bibinfo  {journal} {Physical Review B}\ }\textbf {\bibinfo {volume} {87}},\ \bibinfo {pages} {035401} (\bibinfo {year} {2013})}\BibitemShut {NoStop}%
\bibitem [{\citenamefont {Beenakker}(2013)}]{Beenakker2013}%
  \BibitemOpen
  \bibfield  {author} {\bibinfo {author} {\bibfnamefont {C.}~\bibnamefont {Beenakker}},\ }\href {\doibase 10.1146/annurev-conmatphys-030212-184337} {\bibfield  {journal} {\bibinfo  {journal} {Annual Review of Condensed Matter Physics}\ }\textbf {\bibinfo {volume} {4}},\ \bibinfo {pages} {113} (\bibinfo {year} {2013})}\BibitemShut {NoStop}%
\bibitem [{\citenamefont {Zhou}\ \emph {et~al.}(2013)\citenamefont {Zhou}, \citenamefont {Wu}, \citenamefont {Li}, \citenamefont {He},\ and\ \citenamefont {Kou}}]{Zhou2013}%
  \BibitemOpen
  \bibfield  {author} {\bibinfo {author} {\bibfnamefont {J.}~\bibnamefont {Zhou}}, \bibinfo {author} {\bibfnamefont {Y.-J.}\ \bibnamefont {Wu}}, \bibinfo {author} {\bibfnamefont {R.-W.}\ \bibnamefont {Li}}, \bibinfo {author} {\bibfnamefont {J.}~\bibnamefont {He}}, \ and\ \bibinfo {author} {\bibfnamefont {S.-P.}\ \bibnamefont {Kou}},\ }\href@noop {} {\bibfield  {journal} {\bibinfo  {journal} {Europhysics Letters}\ }\textbf {\bibinfo {volume} {102}},\ \bibinfo {pages} {47005} (\bibinfo {year} {2013})}\BibitemShut {NoStop}%
\bibitem [{\citenamefont {Biswas}(2013)}]{Biswas2013}%
  \BibitemOpen
  \bibfield  {author} {\bibinfo {author} {\bibfnamefont {R.~R.}\ \bibnamefont {Biswas}},\ }\href@noop {} {\bibfield  {journal} {\bibinfo  {journal} {Physical Review Letters}\ }\textbf {\bibinfo {volume} {111}},\ \bibinfo {pages} {136401} (\bibinfo {year} {2013})}\BibitemShut {NoStop}%
\bibitem [{\citenamefont {Sarma}\ \emph {et~al.}(2015)\citenamefont {Sarma}, \citenamefont {Freedman},\ and\ \citenamefont {Nayak}}]{Sarma2015}%
  \BibitemOpen
  \bibfield  {author} {\bibinfo {author} {\bibfnamefont {S.~D.}\ \bibnamefont {Sarma}}, \bibinfo {author} {\bibfnamefont {M.}~\bibnamefont {Freedman}}, \ and\ \bibinfo {author} {\bibfnamefont {C.}~\bibnamefont {Nayak}},\ }\href@noop {} {\bibfield  {journal} {\bibinfo  {journal} {npj Quantum Information}\ }\textbf {\bibinfo {volume} {1}},\ \bibinfo {pages} {1} (\bibinfo {year} {2015})}\BibitemShut {NoStop}%
\bibitem [{\citenamefont {Liu}\ and\ \citenamefont {Franz}(2015)}]{Liu2015}%
  \BibitemOpen
  \bibfield  {author} {\bibinfo {author} {\bibfnamefont {T.}~\bibnamefont {Liu}}\ and\ \bibinfo {author} {\bibfnamefont {M.}~\bibnamefont {Franz}},\ }\href {\doibase 10.1103/PhysRevB.92.134519} {\bibfield  {journal} {\bibinfo  {journal} {Physical Review B}\ }\textbf {\bibinfo {volume} {92}},\ \bibinfo {pages} {134519} (\bibinfo {year} {2015})}\BibitemShut {NoStop}%
\bibitem [{\citenamefont {Murray}\ and\ \citenamefont {Vafek}(2015)}]{Murray2015}%
  \BibitemOpen
  \bibfield  {author} {\bibinfo {author} {\bibfnamefont {J.~M.}\ \bibnamefont {Murray}}\ and\ \bibinfo {author} {\bibfnamefont {O.}~\bibnamefont {Vafek}},\ }\href@noop {} {\bibfield  {journal} {\bibinfo  {journal} {Physical Review B}\ }\textbf {\bibinfo {volume} {92}},\ \bibinfo {pages} {134520} (\bibinfo {year} {2015})}\BibitemShut {NoStop}%
\bibitem [{\citenamefont {Smith}\ \emph {et~al.}(2016)\citenamefont {Smith}, \citenamefont {Tanaka},\ and\ \citenamefont {Nagai}}]{Smith2016}%
  \BibitemOpen
  \bibfield  {author} {\bibinfo {author} {\bibfnamefont {E.~D.}\ \bibnamefont {Smith}}, \bibinfo {author} {\bibfnamefont {K.}~\bibnamefont {Tanaka}}, \ and\ \bibinfo {author} {\bibfnamefont {Y.}~\bibnamefont {Nagai}},\ }\href@noop {} {\bibfield  {journal} {\bibinfo  {journal} {Physical Review B}\ }\textbf {\bibinfo {volume} {94}},\ \bibinfo {pages} {064515} (\bibinfo {year} {2016})}\BibitemShut {NoStop}%
\bibitem [{\citenamefont {Wan}\ \emph {et~al.}(2023{\natexlab{b}})\citenamefont {Wan}, \citenamefont {Zheliuk}, \citenamefont {Yuan}, \citenamefont {Peng}, \citenamefont {Zhang}, \citenamefont {Liang}, \citenamefont {Zeitler}, \citenamefont {Wiedmann}, \citenamefont {Hussey}, \citenamefont {Palstra} \emph {et~al.}}]{Wan2023}%
  \BibitemOpen
  \bibfield  {author} {\bibinfo {author} {\bibfnamefont {P.}~\bibnamefont {Wan}}, \bibinfo {author} {\bibfnamefont {O.}~\bibnamefont {Zheliuk}}, \bibinfo {author} {\bibfnamefont {N.~F.}\ \bibnamefont {Yuan}}, \bibinfo {author} {\bibfnamefont {X.}~\bibnamefont {Peng}}, \bibinfo {author} {\bibfnamefont {L.}~\bibnamefont {Zhang}}, \bibinfo {author} {\bibfnamefont {M.}~\bibnamefont {Liang}}, \bibinfo {author} {\bibfnamefont {U.}~\bibnamefont {Zeitler}}, \bibinfo {author} {\bibfnamefont {S.}~\bibnamefont {Wiedmann}}, \bibinfo {author} {\bibfnamefont {N.~E.}\ \bibnamefont {Hussey}}, \bibinfo {author} {\bibfnamefont {T.~T.}\ \bibnamefont {Palstra}},  \emph {et~al.},\ }\href@noop {} {\bibfield  {journal} {\bibinfo  {journal} {Nature}\ ,\ \bibinfo {pages} {1}} (\bibinfo {year} {2023}{\natexlab{b}})}\BibitemShut {NoStop}%
\end{thebibliography}%

\end{document}